\documentclass[a4paper]{jpconf}
\newcommand{\p}[1]{(\ref{#1})}
\usepackage{graphicx}
\usepackage{amssymb}
\usepackage{amsmath}
\begin{document}
\title{Anisotropic pressure in the quark core of a strongly magnetized
hybrid star}

\author{A A Isayev}

\address{$^1$Kharkov Institute of
Physics and Technology, Academicheskaya Street 1,
 Kharkov, 61108, Ukraine}
\ead{isayev@kipt.kharkov.ua}

\begin{abstract}
The impact of a strong magnetic field, varying with the total baryon
number density, on thermodynamic properties of strange quark matter
(SQM) in the core of a magnetized hybrid star is considered  at zero
temperature within the framework of the Massachusetts Institute of
Technology (MIT) bag model. It is clarified that the central
magnetic field strength is bound from above by the value at which
the derivative of the longitudinal pressure with respect to the
baryon number density vanishes first somewhere in the quark core
under varying the central field. Above this upper bound, the
instability along the magnetic field is developed in magnetized SQM.
The total energy density, longitudinal and transverse pressures are
found as functions of the total baryon number density.
\end{abstract}

\section{Introduction}

The study of the QCD phase diagram under extreme conditions of
temperature or density is continuing to be a hot research issue. The
other  factor which can significantly influence the structure of the
QCD phase diagram is  magnetic field. 
Strong magnetic fields of about $H\sim10^{18}$~G (RHIC), or even of
an order of magnitude larger (LHC), are generated in noncentral
high-energy heavy-ion collisions~\cite{SIT,DH}. The electric charge
separation with respect to the reaction plane of  colliding nuclei
due to the the chiral magnetic effect~\cite{NPA08Kharzeev}  could be
one of the observable imprints of strong magnetic fields generated
in heavy-ion collisions. In the astrophysical context, for a special
class of neutron stars, called magnetars,  the magnetic field
strength at the surface can reach the values of about
$10^{14}$-$10^{15}$~G. Even stronger magnetic fields up to
$10^{19}$--$10^{20}$~G may potentially occur in the inner cores of
magnetars~\cite{BCP}. The possible imprint of such ultrastrong
magnetic fields could be the large pulsar kick velocities as a
result of asymmetric neutrino emission in direct Urca processes in
the dense core of a magnetized neutron star~\cite{ARDM}. The origin
of strong magnetic fields of magnetars is still under discussion,
and, among other possibilities, it is not excluded that this can be
due to
 spontaneous ordering of hadron~\cite{IY,PRC06I,IY3,IY4}, or quark~\cite{TT}
spins in the dense interior of a neutron star.

Thus,  the study of  thermodynamic properties of nuclear matter in a
strong magnetic field is the problem of a considerable interest. In
particular, the pressure anisotropy, exhibited in the difference
between the longitudinal and transverse (along and perpendicular to
the magnetic field) pressures, becomes important for strongly
magnetized matter~\cite{Kh,IY_PRC11,IY_PLB12,JPG13IY,NPA13SMS}. In
this study, we consider  strongly magnetized SQM, composed of
deconfined up, down and strange quarks,
within the framework of the MIT bag model~\cite{FJ,CJJ}.
The relevant astrophysical object is a hybrid star whose quark core
is surrounded with the hadronic crust. It was clarified earlier
that, if the uniform magnetic field exceeds some critical value, the
longitudinal pressure becomes negative resulting in the appearance
of the longitudinal instability in SQM~\cite{JPG13IY}. The value of
the corresponding critical field represents, in fact, the upper
bound on the magnetic field  in the quark core of a hybrid star when
spatial nonuniformity in the field distribution is disregarded. In
this research, we would like to extend the previous
consideration~\cite{JPG13IY} to the case of the
spatially nonuniform magnetic field distribution. 

\section{Numerical results and discussion}

For the details of the formalism one can address to
Ref.~\cite{JPG13IY}.  We use the simplified variant of the MIT bag
model, in which quarks are considered as free fermions moving inside
a finite region of space called a "bag". The effects of the quark
confinement are implemented by introducing the bag pressure $B$.
Note that SQM can represent the ground state of matter, or can be
metastable. In order to be absolutely stable, the energy per baryon
of magnetized SQM should be less than that of the most stable
$^{56}$Fe nucleus under the zero external pressure and temperature.
Further we will be interested in the astrophysical scenario, in
which SQM forms the core of a hybrid star and, hence, is metastable
at zero pressure. The gravitational pressure from the outer hadronic
layers stabilizes quark matter in the core.

In the MIT bag model, the total energy density $E$, the longitudinal
$p_l$ and transverse $p_t$ pressures in magnetized
quark matter  read 
\begin{align}
E&= \Omega+\sum_i\mu_i\varrho_i+\frac{H^2}{8\pi}+B, 
\label{E}\\
p_{l}&=-\Omega-\frac{H^2}{8\pi}-B,\label{pl}\\
p_{t}&=-\Omega-HM+\frac{H^2}{8\pi}-B, \label{pt}
\end{align}
where $M=-\frac{\partial\Omega}{\partial H}$
is the total magnetization,  $\Omega=\sum_i\Omega_i$,  $\Omega_i$
and $\varrho_i=-\frac{\partial\Omega_i}{\partial\mu_i}$
are the
thermodynamic potential   and number density, respectively, for
fermions of $i$th species with the chemical potential $\mu_i$,
including $u,d,s$ quarks plus electrons to ensure charge neutrality
and beta equilibrium with respect to the weak processes 
occurring in the quark core of a hybrid star.

In the previous research~\cite{JPG13IY}, the impact of the uniform
magnetic field on thermodynamic properties of SQM was considered. In
a more realistic study, it is necessary to take into account that
the magnetic field varies from the core to the surface of a star.
Following Ref.~\cite{BCP}, we will model this change by the
dependence of the magnetic field on the baryon density of the form

\begin{align}
 H(\varrho_B)= H_{s}+H_{cen} 
\Bigl(1-e^{-\beta\bigl(\frac{\varrho_B}{\varrho_0}\bigr)^\gamma}\Bigr).
\label{H_ro_exp} 
\end{align}

Here $H_{cen}$ and $H_s$ are the magnetic field strengths in the
center (assuming that the central baryon density is essentially
larger than the nuclear saturation density $\varrho_0=0.16$~${\rm
fm}^{-3}$) and at the surface of a star, respectively; $\beta$ and
$\gamma$ are the model parameters. We set the surface field
$H_s=10^{15}$~G, and the central density is chosen to be
$\varrho_{cen}=7\varrho_0$. In Eq.~\p{H_ro_exp}, we use the model
parameter sets $\beta=0.02,\gamma=3$ from Ref.~\cite{RPPP} (slow
varying magnetic field), and $\beta=0.001,\gamma=6$ from
Ref.~\cite{NPA13SMS} (fast varying magnetic field). In the
subsequent calculations,  the same quark current masses are employed
as in the studies~\cite{JPG13IY,HHRS}, and we adopt
$B=76$~MeV/fm$^3$, which is slightly larger than the upper bound
$B_u\simeq75$~MeV/fm$^3$ from the absolute stability window.

\begin{figure*}[tb]
\begin{center}
\includegraphics[width=0.7\linewidth]{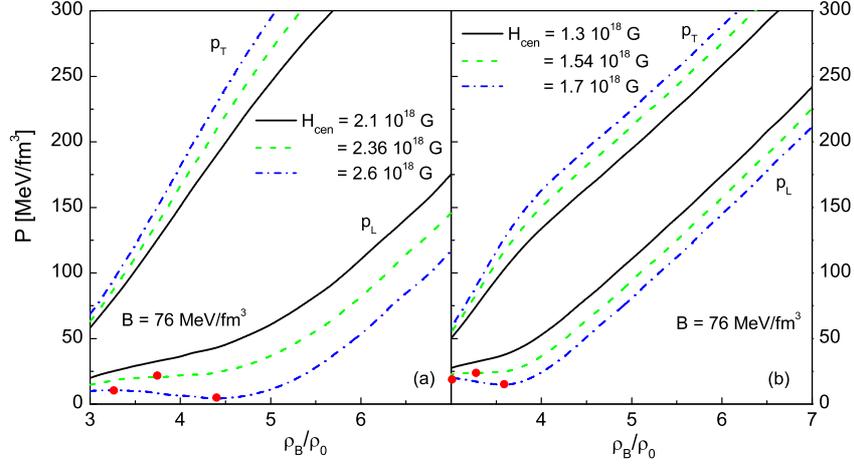}
\end{center}
\vspace{-2ex} \caption{Transverse $p_t$ (three upper curves) and
longitudinal $p_l$ (three lower curves) pressures in magnetized SQM
as functions of the total baryon number density, corresponding to
Eq.~\p{H_ro_exp} with: (a)   $\beta=0.02,\gamma=3$ and (b)
$\beta=0.001,\gamma=6$ for $H_s=10^{15}$~G and variable central
field $H_{cen}$. The full dots correspond to the points where
$p_l\,'(\varrho_B)=0$.}
\label{fig3}\vspace{-0ex}
\end{figure*}

 Further we will assume that the quark phase appears at
the total baryon density $\varrho_B\sim3\varrho_0$~\cite{FW}, and,
hence, the baryon density for magnetized SQM changes in the range
$3\varrho_0\lesssim \varrho_B \leq 7\varrho_0$. The main strategy in
the further calculations of the anisotropic pressure is as follows.
We vary the central magnetic field strength $H_{cen}$ in
Eq.~\p{H_ro_exp},  and like to determine at which baryon density
within the above range the onset of the longitudinal instability
occurs. Fig.~\ref{fig3} shows the transverse $p_t$ and longitudinal
$p_l$ pressures  as functions of the total baryon number density for
the above parametrizations of the magnetic field strength. The
general tendency is that, under increasing the central field
$H_{cen}$, the transverse pressure $p_t$ increases while the
longitudinal pressure $p_l$ decreases. Also, the transverse pressure
always increases with the total baryon density $\varrho_B$, while
the dependence of the longitudinal pressure $p_l$ on $\varrho_B$ can
be different. Let us consider first the case of the slow varying
magnetic field with $\beta=0.02,\gamma=3$ in Eq.~\p{H_ro_exp}
(Fig.~\ref{fig3}a). Under increasing the central field $H_{cen}$,
the longitudinal pressure, at first, remains to be increasing
function of the baryon density with $p_l\,'(\varrho_B)>0$. However,
under further increasing  the central field, the
 curve $p_l(\varrho_B)$ is bending down in its middle part.  At $H_{cen}\approx2.36\cdot 10^{18}$~G,
 the derivative $p_l\,'(\varrho_B)$ vanishes at
 $\varrho_B\approx3.74\varrho_0$
 (the
corresponding point on the curve is marked by the full red dot)
while remaining positive for other baryon densities from the
interval under consideration. Under further increasing the central
magnetic field, there appears the part of the curve characterized by
the negative derivative $p_l\,'(\varrho_B)<0$, contrary to the
stability constraint $p_l\,'(\varrho_B)>0$. Hence, the corresponding
states of magnetized SQM are unstable and the instability is
developed along the magnetic field direction. The onset of
instability corresponds to $H_{cen}\approx2.36\cdot 10^{18}$~G, at
which the derivative $p_l\,'(\varrho_B)$ vanishes first. This value
represents the upper bound on the central magnetic field strength in
the quark core of a hybrid star in the case of the slow varying
magnetic field with $\beta=0.02,\gamma=3$ in Eq.~\p{H_ro_exp}. For
the fast varying field with $\beta=0.001,\gamma=6$, 
the upper bound on the central magnetic field strength is somewhat
smaller, $H_{cen}\approx1.54\cdot 10^{18}$~G, for which the
derivative $p_l\,'(\varrho_B)$ vanishes at
$\varrho_B\approx3.27\varrho_0$ (cf. Fig.~\ref{fig3}b).

\begin{figure*}[tb]
\begin{center}
\includegraphics[width=0.75\linewidth]{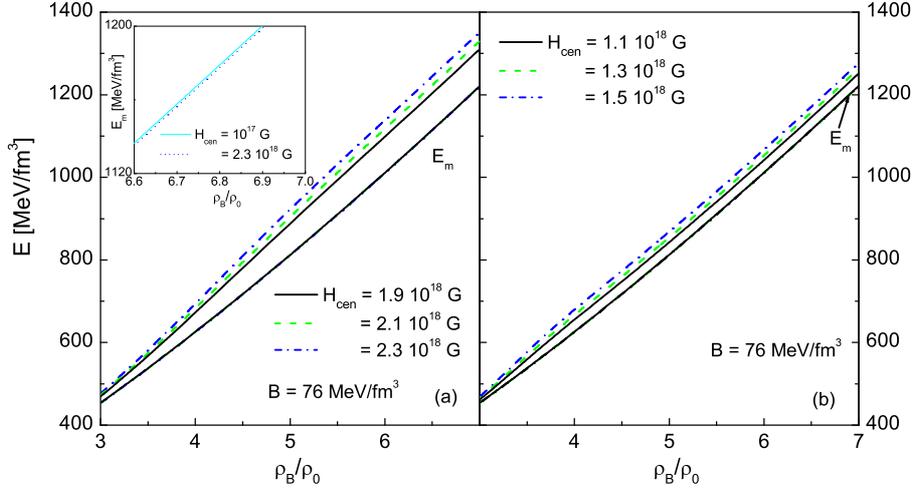}
\end{center}
\vspace{-2ex} \caption{The energy density $E$ (upper curves apart
from the most lower curve) and the matter energy density $E_m$ (the
most lower curve) of magnetized strange quark matter vs. the total
baryon number density for the same parametrizations of the magnetic
field strength as in
Fig.~\ref{fig3}  and variable central field $H_{cen}$. 
The insert in the left panel shows the matter energy density $E_m$
in the high baryon density region.  } \label{fig4}\vspace{-0ex}
\end{figure*}

Fig.~\ref{fig4} shows the energy density $E$ of the system  and its
matter part  $E_m\equiv E-E_f$ ($E_f=\frac{H^2}{8\pi}$  being the
magnetic field energy density) as functions of the total baryon
number density for the above parametrizations   of the magnetic
field strength. The curves for the matter part are practically
indistinguishable for the different values of the central magnetic
field, used in calculations with each parameter set, and look almost
as one curve. This figure allows to estimate the relative role of
the matter $E_m$ and magnetic field $E_f$ contributions to the total
energy density $E$. It is seen that the matter part dominates over
the field part at such baryon number densities and magnetic field
strengths $H_s, H_{cen}$. Also, the insert in the left panel shows
the matter part
 $E_m$ as a function of the baryon density for two differing by an
order of magnitude values of the central field $H_{cen}$ in order to
demonstrate the effect of Landau diamagnetism. It is seen that the
matter part  is smaller in the case of the stronger central magnetic
field. Nevertheless, the overall effect of the stronger central
field is to increase the total energy of the system because of the
pure magnetic field  contribution.

\begin{figure*}[tb]
\begin{center}
\includegraphics[height=10.2cm,keepaspectratio]{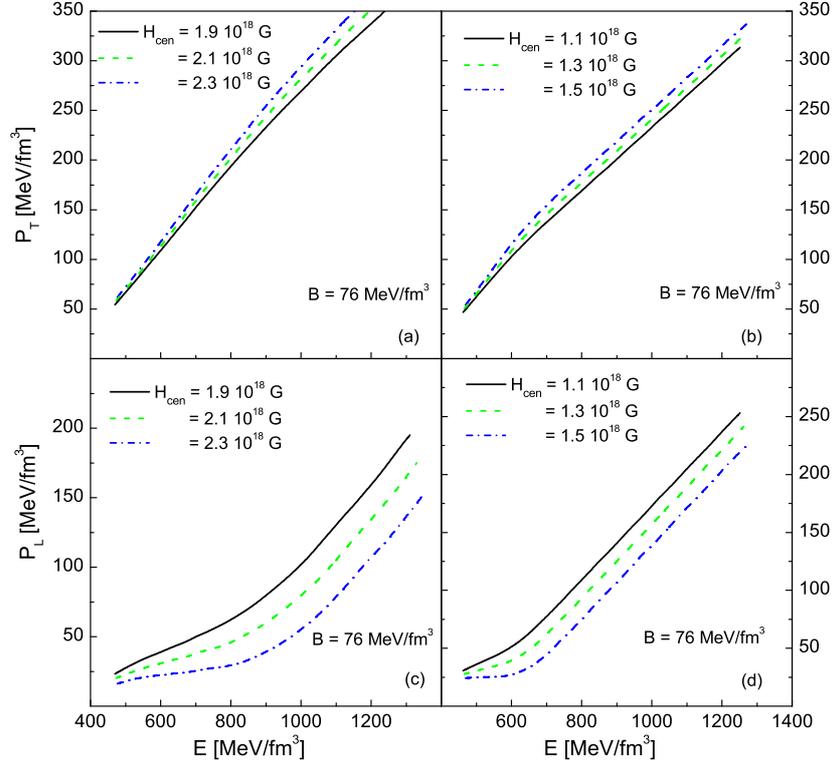}
\end{center}
\vspace{-2ex} \caption{The transverse $p_t$ (panels (a), (b)) and
longitudinal $p_l$ (panels (c), (d)) pressures in magnetized strange
quark matter as functions of the total energy density $E$ for the
same parametrizations of the magnetic field strength as in
Fig.~\ref{fig3}  and variable central field $H_{cen}$.
  } \label{fig5}\vspace{-0ex}
\end{figure*}

In   strong magnetic fields beyond some threshold value $H> H_{th}$,
with $10^{17}< H_{th}\lesssim10^{18}$~G, the total pressure in
magnetized strange quark matter becomes essentially
anisotropic~\cite{JPG13IY}. Hence, the equation of state (EoS) of
the system becomes also highly anisotropic. Fig.~\ref{fig5}, showing
the dependence of the transverse $p_t$ (upper row) and longitudinal
$p_l$ (lower row) pressures on the energy density $E$ of magnetized
strange quark matter, explicitly demonstrates this moment. Because
in all cases we choose the central field smaller than the
corresponding upper bound, all quantities $p_t$, $p_l$ and $E$ are
the increasing functions of the baryon density, and, hence, $p_t(E)$
and $p_l(E)$ are also the increasing functions.

In summary, we have considered the impact of varying with the total
baryon number density magnetic field  on thermodynamic properties of
SQM at zero temperature under conditions relevant to the cores of
strongly magnetized hybrid stars.  The total energy density $E$ of
magnetized SQM, transverse $p_t$
 and  longitudinal $p_l$ pressures have been calculated as functions of the total baryon
number density. Also,  the highly anisotropic EoS has been
determined in the form of $p_t(E)$ and $p_l(E)$ dependences. It has
been clarified that the central magnetic field in the core of a
hybrid star is bound from above by the value, at which  the
derivative of the longitudinal pressure $p_l\,'(\varrho_B)$ vanishes
first under varying the central field. Above this upper bound,  the
instability along the magnetic field direction is developed in
magnetized SQM.

In this study, we have considered the thermodynamic properties of
strongly magnetized SQM at zero temperature. It would be of interest
also to extend this research to finite
temperatures~\cite{IY_PRC11,DMS}, when new interesting effects could
occur, such as the unusual behavior of the entropy of a spin
polarized state~\cite{PRC05I,PRC07I,IY10}.

\end{document}